\documentclass[runningheads]{llncs}
\usepackage{bbding}
\usepackage{graphicx}
\usepackage{hyperref}
\usepackage{booktabs}
\usepackage{tipa}
\begin{document}
\renewcommand{\arraystretch}{1.5}
\setlength{\tabcolsep}{2 pt}
\title{Enhancing MRI Brain Tumor Segmentation with an Additional Classification Network\thanks{This work was done when Hieu Nguyen and Tung Le were AI Interns at Medical Imaging Department, Vingroup Big Data Institute (VinBigdata).}}
\titlerunning{Enhancing MRI Brain Tumor Segmentation by Classification}
\author{Hieu T. Nguyen\inst{1,2,*,}$ \textrm{\Envelope}$ \and %$^\Envelope$^{,\dag}$ \and
Tung T. Le\inst{1,3,*} %$^{,\dag}$  \and
Thang V. Nguyen\inst{1} \and 
Nhan T. Nguyen\inst{1}}
\authorrunning{Nguyen et al.}
\institute{Medical Imaging Department, Vingroup Big Data Institute (VinBigdata) \and Hanoi University of Science and Technology (HUST) \and University of Engineering and Technology (UET), VNU\\
${\textrm{\Envelope}}$ Correspondence to Hieu Nguyen; e-mail: \email{hieu.nt170073@sis.hust.edu.vn}\\
$^*$ These authors share first authorship on this work. } 
\maketitle
\begin{abstract}
Brain tumor segmentation plays an essential role in medical image analysis. In recent studies, deep convolution neural networks (DCNNs) are extremely powerful to tackle tumor segmentation tasks. We propose in this paper a novel training method that enhances the segmentation results by adding an additional classification branch to the network. The whole network was trained end-to-end on the Multimodal Brain Tumor Segmentation Challenge (BraTS) 2020 training dataset. 
% On the BraTS's validation set, it achieved an average Dice score of $78.43\%$, $89.99\%$ and $84.22\%$ 
On the BraTS's test set, it achieved an average Dice score of $80.57\%$, $85.67\%$ and $82.00\%$, as well as Hausdorff distances $(95\%)$ of $14.22$, $7.36$ and $23.27$, 
respectively for the enhancing tumor, the whole tumor and the tumor core.

\end{abstract}
\section{Introduction}

Gliomas are the most common primary brain malignancies, with different degrees of aggressiveness, variable prognosis and various heterogeneous histological sub-regions~\cite{bakas2017advancing,bakas2017segmentationLGG,bakas2017segmentationGBM}. One objective of The Brain Tumor segmentation (BraTS) challenge is to identify state-of-the-art machine learning methods for segmentation of brain tumors in magnetic resonance imaging (MRI) scans \cite{menze2014multimodal,bakas2018identifying}. One MRI data sample consists of a native T1-weighted scan (T1), a post-contrast T1-weighted scan (T1Gd), a native T2-weighted scan (T2), and a T2 Fluid Attenuated Inversion Recovery (T2-FLAIR) scan. However, each tumor-region-of-interest (TRoI) is visible in one pulse. Specifically, the whole tumor is visible in T2-FLAIR, the tumor core is visible in T2, and the enhancing tumor is visible in T1Gd.

An accurate deep learning segmentation model not only can save time for neuroradiologists but provides a reliable result for further tumor analysis. Recently, deep learning approaches have consistently surpassed traditional computer vision methods~\cite{havaei2017brain,kamnitsas2017efficient,pereira2016brain,shen2017boundary,zhao2018deep}. Specifically, convolutional neural networks (CNN) are able to learn deep representative features to generate accurate segmentation mask both in 2D and 3D medical images. 

The BraTS 2020 training dataset, which comprises 369 cases for training and 125 cases for validation, is manually annotated by both clinicians and board-certified radiologists. Each tumor is segmented into enhancing tumor, peritumoral edema, and the necrotic and non-enhancing tumor core. To evaluate the segmentation performance, various metrics are used: Dice score, Hausdorff distance (95\%), sensitivity and specificity. 

Since the introduction of U-Net~\cite{ronneberger2015unet} in 2015, various types of U-shape DCNN have been proposed and gained significant results in medical image segmentation tasks. In BraTS 2017, Kamnitsas et al.~\cite{kamnitsas2017ensembles}, who was the winner of the segmentation challenge, explored Ensembles of Multiple Models and Architecture (EMMA) for robust performance by combining several DCNNs including DeepMedic~\cite{kamnitsas2017efficient}, 3D FCN~\cite{long2015fully} and 3D U-Net~\cite{cciccek20163d}. In BraTS 2018, Myronenko~\cite{myronenko20183d}, who won segmentation track, utilized asymmetrically large encoder to extract deep image features, and the decoder part reconstructs dense segmentation masks. The authors also added the variational autoencoder (VAE) branch in order to regularize the network. In BraTS 2019, Jiang et al.\cite{jiang2019two}, who recently achieved the highest score on private test set, deployed two-stage cascaded U-Net which basically stacked 2 U-Net networks together. In the first stage, they used a variant of U-Net to train a coarse prediction. In the next stage, they increased the network capacity by using 2 decoders simultaneously. The model was trained in an end-to-end manner and achieved the best result.

\textbf{Contribution.} Through exploratory model analysis after training, we notice that deep learning segmentation models sometimes make false positive predictions. To bridge the gap between segmentation model efficiency and avoid these problems, we proposed a novel end-to-end training method by combining both segmentation and classification. The classification branch helps to predict whether a mask slice contains region of interest as well as to regularize the segmentation branch. We explored this approach with 2 architectures which are variant of nested U-Net~\cite{zhou2018unet++} and Bi-directional Feature Pyramid Network (BiFPN) \cite{tan2019efficientdet}. Our method achieved Dice score of $80.57\%$, $85.67\%$ and $82.00\%$ respectively for the enhancing tumor, the whole tumor and the tumor core on the test dataset of the 2020 BraTS challenge\footnote{\url{https://www.med.upenn.edu/cbica/brats2020/data.html}}.

\section{Methods}
In this section, we describe the proposed approach in which two different models, BiFPN and Nested U-Net, are leveraged as the base segmentation architecture, enhanced by a classifier head. While the segmentation head largely relies on local features to segment tumor area, the classification branch leverages global features of the whole slice as well as neighbors slices to aid segmentation task. The main advantage of classification head is that it significantly reduces false positive regions since minute, high intensity regions of enhancing tumor are often confused with other non-tumor, high intensity regions. In addition, to tackle small batch size problem when using batch-norm, we deploy Group Normalization~\cite{wu2018group} with number groups of 8 instead. 

\subsection{Bi-directional feature pyramid network}
\begin{figure}
\centering
\includegraphics[width=\textwidth]{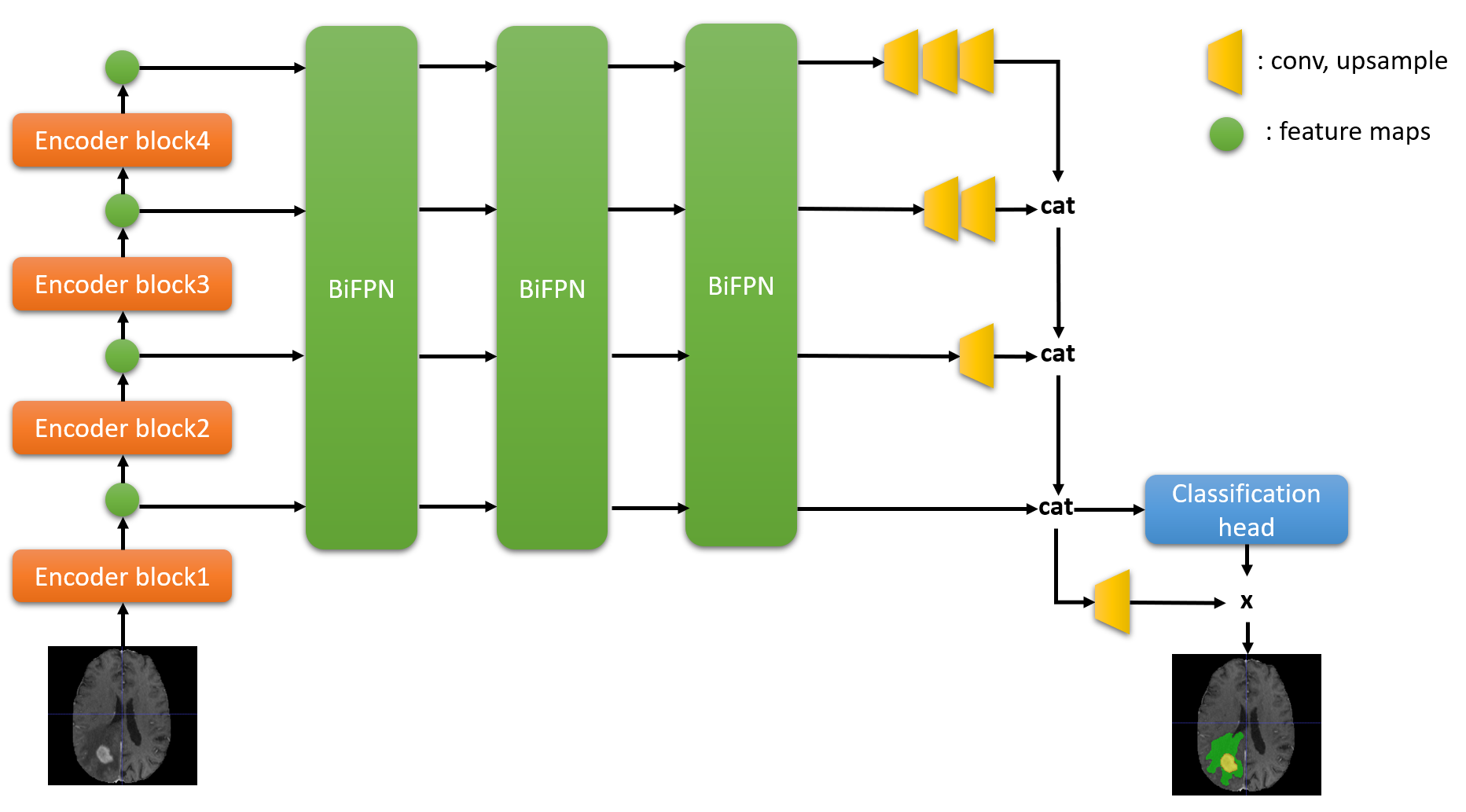}
\caption{Overview of the BiFPN architecture with Classifier.} \label{meta_arch}
\end{figure}

In this approach, an encoder-decoder based network is leveraged \cite{zhou2018unet++,kirillov2019panoptic} with an additional classification branch to further enhance segmentation results. The classification head is placed at the end of the encoder to classify whether an image slice has tumor region. In the following subsections, we describe the details of the encoder and decoder parts (see Fig.\ref{meta_arch}). 

\subsubsection{Encoder}
\begin{figure}
\centering
\includegraphics[width=\textwidth]{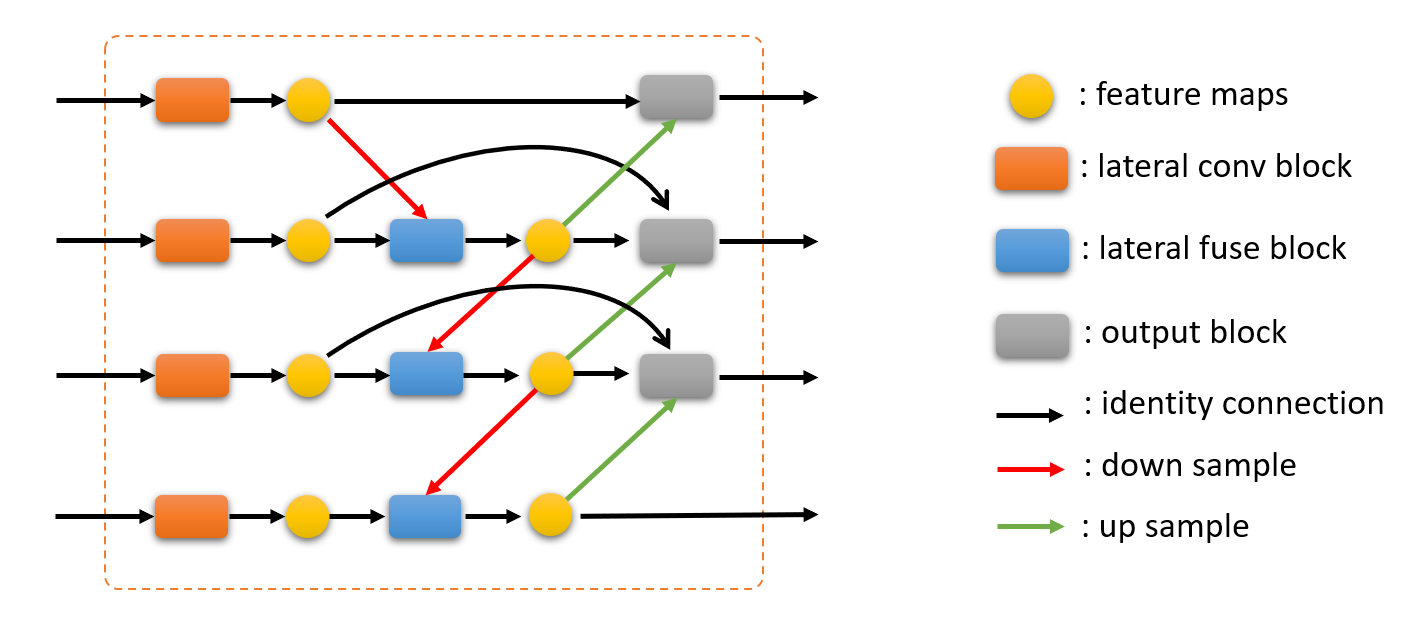}
\caption{Illustration of the BiFPN layer.} \label{bifpn}
\end{figure}

    For the encoder part, we exploited residual block \cite{he2016deep} for features extraction with the number of channels being doubled after each convolutional layer of stride 2, which results in multi-scale features maps for the latter part. There are four scales of feature maps, where the smallest one were 16 times smaller than the input image (see Table.\ref{resblock} for the details of the feature extractor). In order to combine features of multiple sizes, we adapted the BiFPN layer from EfficientDet architecture \cite{tan2019efficientdet} (see Fig.\ref{bifpn}), which was an improved version of the Feature Pyramid networks \cite{lin2017feature}. We used three consecutive BiFPN layers with feature dimensions of 256, as deeper networks did not improve performance. 

\begin{table}
\caption{Details of the feature extractor, where \textit{conv3} is $3\times3\times3$ convolution and GN denotes group norm. Note that output shape of encoder blocks correspond to input image of shape $4\times128\times128\times96$}\label{resblock}

\centering
%\begin{table}[]
\scriptsize{
\begin{tabular}{l|l|c|l}
\hline
\hline
Block                       & Details                                                                                                                            & Repeat & Output size   \\ \specialrule{.1em}{.1em}{.1em} 
Encoder block1              & \begin{tabular}[c]{@{}l@{}}(conv3 stride2, GN, dropout, ReLU, \\ conv3 stride1, GN, dropout, ReLU)\\ + conv3 stride2\end{tabular} & 1      & $16 \times 64 \times 64 \times 48$ \\ \hline
Encoder block2              & \begin{tabular}[c]{@{}l@{}}(conv3 stride2, GN, dropout, ReLU, \\ conv3 stride1, GN, dropout, ReLU)\\ + conv3 stride2\end{tabular} & 1      & $32 \times 32 \times 32 \times 24$ \\ \hline
Encoder block3              & \begin{tabular}[c]{@{}l@{}}(conv3 stride2, GN, dropout, ReLU, \\ conv3 stride1, GN, dropout, ReLU)\\ + conv3 stride2\end{tabular} & 1      & $64 \times 16 \times 16 \times 12$ \\ \hline
Encoder block4              & \begin{tabular}[c]{@{}l@{}}(conv3 stride2, GN, dropout, ReLU, \\ conv3 stride1, GN, dropout, ReLU)\\ + conv3 stride2\end{tabular} & 1      & $128 \times 8 \times 8 \times 6$   \\ \hline \hline
\end{tabular}
}
% \end{table}

% # begin, end tabular
\end{table}

\subsubsection{Decoder}
    In the decoder part, we followed the design of semantic segmentation branch of the Panoptic Feature Pyramid networks \cite{kirillov2019panoptic}. Each feature map from the BiFPN layers was put into a number of up-sample blocks, depending on the spatial size. Each up-sample block consists of a 3$\times$3$\times$3 convolution, group-norm and ReLU, followed by 2$\times$ trilinear interpolation, and the feature dimension is fixed to 256. Due to GPU memory constraint, all feature maps were up-sampled to a common size, which is half of the input image size, then were concatenated before putting into the final upsample block, which has a 1$\times$1$\times$1 convolution with 3 filters corresponding to three classes of tumor regions, and subsequently a 2$\times$ trilinear interpolation layer.

\subsection{Nested U-Net}
\begin{figure}
\includegraphics[width=\textwidth]{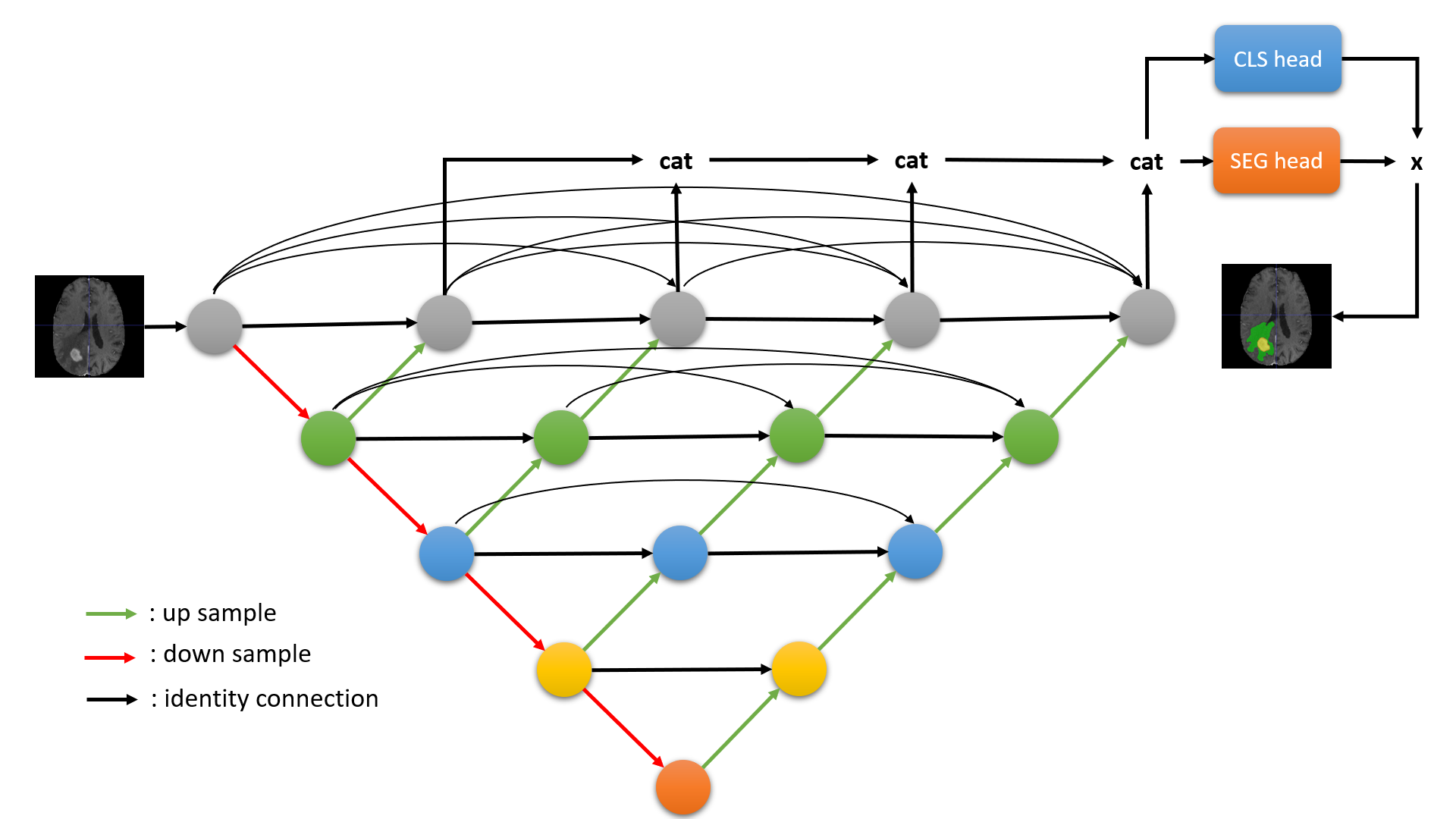}
\caption{Overview of the nested UNET architecture with an additional Classifier.} \label{nested_unet}
\end{figure}

\subsubsection{Skip pathways}
According to Zhou et al.~\cite{zhou2018unet++}, nested U-Net (UNet++) proposed dense convolution block whose number of convolution layers depend on the pyramid level. Therefore, they re-designed skip pathways to bring the semantic level of the encoder feature maps closer to that of the feature maps awaiting in the decoder. Fig.~\ref{nested_unet} clarifies how the feature maps travel through the top skip pathway of UNet++.

\subsubsection{Deep supervision}
In order to take advantage of lower feature maps, we used deep supervision~\cite{lee2015deeply} wherein the final segmentation map was selected from all segmentation branches averaged. Instead of using another layer before upsampling to the map size, our final prediction mask upsamples directly from last layer. Each layer contains ReLU activation, followed by 2 convolution layers.

\subsection{Classification branch}
    In all segmentation architectures, the concatenated feature maps before the final blocks were used as the input for the classification head. While the feature maps for classifier in UNET++ have the same spatial size as the input image, its counterpart in BiFPN are only half of the input, leading to an additional upsample layer in the classification branch of the BiFPN.   
    The classification branch includes: (1) a 3$\times$3$\times$3 convolution block to reduce feature channels; (2) a global average pooling layer that averages feature maps over frontal and sagittal axes to produce slice-wise feature maps along axial axis; (3) a transpose convolution block which is used to upsample axial axis of the feature maps to match input image size (for BiFPN only); (4) a sequence of several BiLSTM layers \cite{hochreiter1997long} which leverages inter-slice dependence from both directions; and (5) a final fully connected layer to classify whether each slice has regions of interested classes. (see Table~\ref{bifpn_cls_head_table} and Table~\ref{unet_cls_head_table})

\begin{table}
\caption{Details of the classification branch in BiFPN, where \textit{conv1d1} is 1-D convolution with kernel size of 1, \textit{conv3d3} is 3-D convolution with kernel size of 3$\times$3$\times$3, \textit{tconv3d3} is 3-D transpose convolution with kernel size of $3 \times 3 \times 3$, GN denotes group-norm. Here output shape of each layer corresponding to input features of shape $1024 \times 64 \times 64 \times 48$.}\label{bifpn_cls_head_table}

\centering
\scriptsize{
% \begin{table}[]
\begin{tabular}{l|l|c|l}
\hline
\hline
Names           & Details                & Repeat & Output size    \\ \specialrule{.1em}{.1em}{.1em}  %\hline
Conv block      & conv3d3, GN, ReLU      & 1      & $512 \times 64 \times 64 \times 48$ \\ \hline
Pool            & global average pooling & 1      & $512 \times 48$       \\ \hline
TransConv block & tconv3d3, GN, ReLU     & 1      & $512 \times 96$      \\ \hline
BiLSTM          & Bi-LSTM with dropout    & 2      & $1024 \times 96$      \\ \hline
FC              & conv1d1                & 1      & $3 \times 96$        \\ \hline \hline
\end{tabular}}
% \end{table}
% # begin, end tabular
\end{table}

\begin{table}
\caption{Details of the classification branch in UNET++. Here output shape of each layer corresponding to input features of shape $128 \times 128 \times 128 \times 128$.}\label{unet_cls_head_table}

\centering
\scriptsize{
% \begin{table}[]
\begin{tabular}{l|l|c|l}
\hline
\hline
Names           & Details                & Repeat & Output size    \\ \specialrule{.1em}{.1em}{.1em}  %\hline
Conv block      & conv3d3, BN, ReLU      & 1      & $256 \times 128 \times 128 \times 128$ \\ \hline
Pool            & global average pooling & 1      & $256 \times 128$       \\ \hline
BiLSTM          & Bi-LSTM with dropout    & 3      & $512 \times 128$      \\ \hline
FC              & conv1d1                & 1      & $3 \times 128$        \\ \hline \hline
\end{tabular}}
% \end{table}
% # begin, end tabular
\end{table}

\subsection{Losses}
\subsubsection{Segmentation loss}
\paragraph{\textbf{Dice loss}}
Dice loss originates from Sørensen–Dice coefficient, which is a statistic developed in 1940s to gauge the similarity between two samples. It was brought in V-Net paper~\cite{milletari2016v}. The Dice Similarity Coefficient (DSC) measures the the degree of overlap between the prediction map and ground truth based on dice coefficient, which is a quantity ranging between 0 and 1 which we aim to maximize. The Dice loss is calculated as
\begin{equation}
\mathcal{L}_{dice} = 1 - \frac{2\sum\limits_{i}^{N}p_{i}g_{i}}{\sum\limits_{i}^{N}p_{i}^{2} + \sum\limits_{i}^{N}g_{i}^{2} + \epsilon },
\end{equation}
where $p_{i}$ is predicted voxels, and $g_{i}$ is ground truth. The sums run over N voxels and we add a small $\epsilon = 1e - 5$ to avoid zero denominator.

\paragraph{\textbf{Focal loss}}
To deal with large class imbalance in the segmentation problem, we also used focal loss \cite{lin2017focal} to penalize more on wrongly segmented regions
\begin{equation}
\mathcal{L}_{focal}(p_{t}) = -\alpha_{t}(1-p_{t})^\gamma \log(p_{t}).
\end{equation}
\indent We directly optimized the label regions (whole tumor, tumor core, and enhancing tumor) with the losses. 
\subsubsection{Classification loss} For the classification branch, we used \textbf{focal loss} and standard \textbf{binary cross entropy loss}. \\

\indent Finally, we summed over all the losses to obtain the final loss.
\begin{equation}
    \mathcal{L}_{total} = \mathcal{L}_{focal\_seg} + \mathcal{L}_{dice} + \mathcal{L}_{focal\_cls} + \mathcal{L}_{BCE}.
\end{equation}

\section{Experiments}
\subsection{Data Pre-processing and Augmentation}
\begin{figure}
\centering
\includegraphics[width=0.5\textwidth]{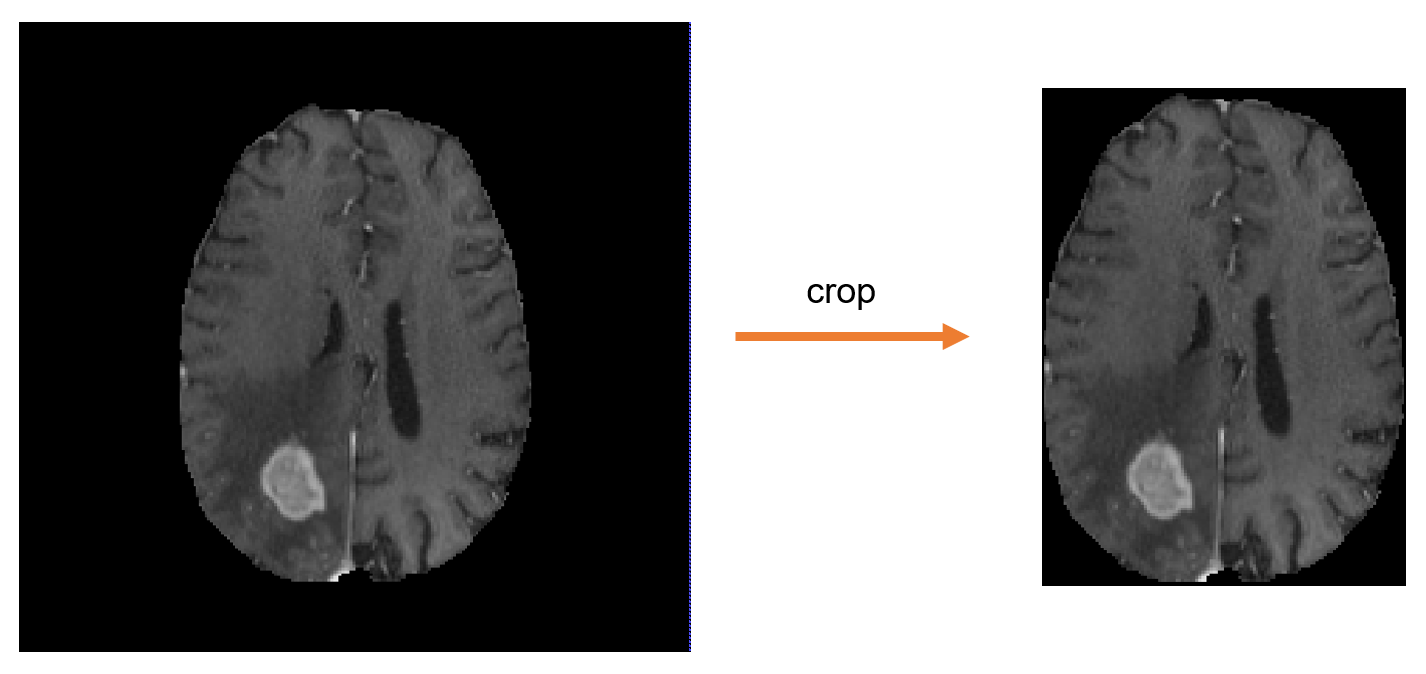}
\caption{Crop out zero-intensity region.} \label{crop}
\end{figure}

%crop, normalize intensity 
For preprocessing, we cropped out zero-intensity regions in order to reduce the image size as well as discard the out-of-interest regions in training (see Figure~\ref{crop}).
To prevent the network from overfitting, we executed several types of data augmentation. Firstly, we applied random flip with probability of 0.5 in every spatial axis. Then, we applied a random scale intensity shift of input in the range of $[0.9,1.1]$. Finally, we applied random intensity shift of input between offset $[-0.1,0.1]$. Both these shift augmentations were applied with the probability of 0.8.
We also applied random crop data with the size of $128 \times 128 \times 128$ and $128 \times 128 \times 96$ voxels due to memory limitation. 
% Through some experiments, we observed that the size 128 works better in our validation set.

\subsection{Training details}
Thanks to Pytorch 1.6, we took advantage of automatic mixed precision to save GPU memory. Adam optimizer was exploited to update model's parameters. Moreover, instead of using fixed learning rate during training, we deployed 2 learning rate schedulers (see below) which were cosine learning rate scheduler~\cite{loshchilov2016sgdr} and polynomial learning rate scheduler~\cite{liu2015parsenet}. Both of the two settings have achieved the same level of performance after 200 epochs with base learning rate of $1e - 3$ and 10 epochs of warming up.

\textbf{Cosine learning rate scheduler} (see Eq.~\ref{cosinelr}) Ignoring the warmup stage and assuming the total number of batches is $T$, initial learning rate $\eta$, the learning rate $\eta_{t}$ at batch $t$ is computed as
\begin{equation}
\label{cosinelr}
    \eta_{t} = \frac {1}{2}(1+\cos (\frac{t\pi}{T}))\eta.
\end{equation}

\textbf{Polynomial learning rate scheduler} (see Eq.~\ref{polynomial}) Ignoring the warm-up stage, $\eta_{0} = 1e - 3$ is initial learning rate, $e$ is epoch counter, $N_{e}$ is total number of epochs, the learning rate $\eta_{t}$ at batch $t$ is computed as
\begin{equation}
\label{polynomial}
    \eta_{t} = \eta_{0} \times (1 - \frac{e}{N_{e}})^{0.9}.
\end{equation}

The cosine learning rate scheduler was used to train in BiFPN model and the polynomial learning rate scheduler was used in nested Unet model. Each network was trained from scratch without neither pretrained weight nor external data on two NVIDIA Tesla V100 32GB RAM. While BiFPN takes a batch of 4 samples, each of shape $4 \times 128 \times 128 \times 96$, UNET++ takes a batch of 2 samples with size of $4 \times 128 \times 128 \times 128$.

\subsection{Inference}
To achieve a robust prediction, we applied test time augmentations (TTA) for every model before averaging them. The augmentations were different flipping in each axis. Finally, we averaged the output before put them through sigmoid function. Decision threshold for both classification and segmentation of three classes was set at 0.5. The negative predictions at slice level of the classification head were used to exclude predicted segmentation regions. While removal of small tumor regions is a simpler approach to address false positive predictions, this method is highly sensitive to volume threshold and can mistakenly exclude the actual tumor which is minute. The classification head provides a more robust solution which can significantly reduce false positive rate, at the same time less likely to miss tiny tumor. 

\section{Results}
    In this section we show the performance of the two architectures with and without the classification head (see Table\ref{val_results}) in terms of Dice score and Hausdorff distance (95\%). The results indicate that classification head improves both architectures by a significant margin. Ensemble of 5 models of the same architecture trained on 5-folds cross-validation gives marginal improvement, while ensemble of the two architectures gives more considerable enhancement.  The results of ensemble of 10 models on the testing set are given in Table \ref{test_results}

\begin{table}
\caption{Mean Dice score and Hausdorff distance of the proposed method on BraTS 2020 validation set.}\label{val_results}
\centering
\scriptsize{

\begin{tabular}{l|r|r|r|r|r|r}
\hline
\hline
Method                            & \multicolumn{3}{l|}{Dice score} & \multicolumn{3}{l}{Hausdorff distance (95\%)} \\ \hline
Validation                        & ET    & WT    & TC     & ET        & WT        & TC          \\ \specialrule{.1em}{.1em}{.1em}
best single Unet++  w/o cls       & 0.7029                 & 0.8967                 & 0.8239              & 42.2474                & 7.4907                 & 9.0179                        \\ \hline
best single Unet++ w. cls         & 0.7742                 & 0.8940                 & 0.8241             & 35.4246                & 8.4361                 & 10.4074                     \\ \hline
ensemble of 5-fold Unet++ w/o cls & 0.7017                 & 0.8953                 & 0.8239            & 47.1436                & 5.8179                 & 11.0075                        \\ \hline
ensemble of 5-fold Unet++ w. cls  & 0.7841                 & 0.8960                 & 0.8233             & 35.4841                & 5.0862                 & 10.0780                   \\ \hline
best single BiFPN w/o cls         & 0.7480                 & 0.8896                 & 0.8400          & 31.1209                & 5.8924                 & 6.9682                        \\ \hline
best single BiFPN w. cls          & 0.7729                 & 0.8881                 & 0.8373             & 21.5720                & 6.9531                 & 6.5573                    \\ \hline
ensemble of 5-fold BiFPN w/o cls  & 0.7471                 & 0.8915                 & 0.8371              & 28.9473                & 5.9362                 & 6.8706                    \\ \hline
ensemble of 5-fold BiFPN w. cls   & 0.7774                 & 0.8914                 & 0.8380           & 24.6944                & 5.9834                 & 6.8527                     \\ \hline
ensemble of 10 models             & \textbf{0.7843}                 & \textbf{0.8999}                 & \textbf{0.8422 }          & \textbf{24.0235 }               & \textbf{5.6808}                 & \textbf{9.5663}                     \\ \hline
\hline
\end{tabular}
}
% \end{table}
% # begin, end tabular
\end{table}

\begin{table}
\caption{Mean Dice score and Hausdorff distance of the proposed method on BraTS 2020 testing set.}\label{test_results}
\centering
\scriptsize{

\begin{tabular}{l|r|r|r|r|r|r}
\hline
\hline
Method                            & \multicolumn{3}{l|}{Dice score} & \multicolumn{3}{l}{Hausdorff distance (95\%)} \\ \hline
Testing                        & ET    & WT    & TC     & ET        & WT        & TC          \\ \specialrule{.1em}{.1em}{.1em}

ensemble of 10 models             & 0.80569                 & 0.85671                & 0.81997           & 14.21938                & 7.35549                 & 23.27358                     \\ \hline
\hline
\end{tabular}
}
% \end{table}
% # begin, end tabular
\end{table}

\section{Conclusion}
In this work, we described a novel training method for brain tumor segmentation from multimodal 3D MRIs. Our results on BraTS 2020 indicated that our model is able to achieve an extremely competitive segmentation result. On the BraTS 2020 test set, the proposed method obtained an average Dice score of $80.57\%$, $85.67\%$ and $82.00\%$, as well as Hausdorff distances $(95\%)$ of $14.22\%$, $7.36\%$ and $23.27\%$, for the enhancing tumor, the whole tumor and the tumor core. In the future, we plan to focus on investigation of new methods for improving small region segmentation as well as classification performance of the network.

\textbf{Acknowledgments} This work was highly supported by Medical Imaging Department at Vingroup Big Data Institute (VinBigdata).
\bibliographystyle{splncs04}
\bibliography{ref.bib}

\begin{thebibliography}{10}
\providecommand{\url}[1]{\texttt{#1}}
\providecommand{\urlprefix}{URL }
\providecommand{\doi}[1]{https://doi.org/#1}

\bibitem{bakas2017segmentationLGG}
Bakas, S., Akbari, H., Sotiras, A., Bilello, M., Rozycki, M., Kirby, J.,
  Freymann, J., Farahani, K., Davatzikos, C.: Segmentation labels and radiomic
  features for the pre-operative scans of the tcga-lgg collection. The cancer
  imaging archive  \textbf{286} (2017)

\bibitem{bakas2017segmentationGBM}
Bakas, S., Akbari, H., Sotiras, A., Bilello, M., Rozycki, M., Kirby, J.,
  Freymann, J., Farahani, K., Davatzikos, C.: Segmentation labels and radiomic
  features for the pre-operative scans of the tcga-gbm collection. the cancer
  imaging archive. Nat Sci Data  \textbf{4},  170117 (2017)

\bibitem{bakas2017advancing}
Bakas, S., Akbari, H., Sotiras, A., Bilello, M., Rozycki, M., Kirby, J.S.,
  Freymann, J.B., Farahani, K., Davatzikos, C.: Advancing the cancer genome
  atlas glioma mri collections with expert segmentation labels and radiomic
  features. Scientific data  \textbf{4},  170117 (2017)

\bibitem{bakas2018identifying}
Bakas, S., Reyes, M., Jakab, A., Bauer, S., Rempfler, M., Crimi, A., Shinohara,
  R.T., Berger, C., Ha, S.M., Rozycki, M., et~al.: Identifying the best machine
  learning algorithms for brain tumor segmentation, progression assessment, and
  overall survival prediction in the brats challenge. arXiv preprint
  arXiv:1811.02629  (2018)

\bibitem{cciccek20163d}
{\c{C}}i{\c{c}}ek, {\"O}., Abdulkadir, A., Lienkamp, S.S., Brox, T.,
  Ronneberger, O.: 3d u-net: learning dense volumetric segmentation from sparse
  annotation. In: International conference on medical image computing and
  computer-assisted intervention. pp. 424--432. Springer (2016)

\bibitem{havaei2017brain}
Havaei, M., Davy, A., Warde-Farley, D., Biard, A., Courville, A., Bengio, Y.,
  Pal, C., Jodoin, P.M., Larochelle, H.: Brain tumor segmentation with deep
  neural networks. Medical image analysis  \textbf{35},  18--31 (2017)

\bibitem{he2016deep}
He, K., Zhang, X., Ren, S., Sun, J.: Deep residual learning for image
  recognition. In: IEEE CVPR. pp. 770--778 (2016).
  \doi{https://doi.org/10.1109/CVPR.2016.90}

\bibitem{hochreiter1997long}
Hochreiter, S., Schmidhuber, J.: Long short-term memory. Neural computation
  \textbf{9}(8),  1735--1780 (1997)

\bibitem{jiang2019two}
Jiang, Z., Ding, C., Liu, M., Tao, D.: Two-stage cascaded u-net: 1st place
  solution to brats challenge 2019 segmentation task. In: International MICCAI
  Brainlesion Workshop. pp. 231--241. Springer (2019)

\bibitem{kamnitsas2017ensembles}
Kamnitsas, K., Bai, W., Ferrante, E., McDonagh, S., Sinclair, M., Pawlowski,
  N., Rajchl, M., Lee, M., Kainz, B., Rueckert, D., et~al.: Ensembles of
  multiple models and architectures for robust brain tumour segmentation. In:
  International MICCAI Brainlesion Workshop. pp. 450--462. Springer (2017)

\bibitem{kamnitsas2017efficient}
Kamnitsas, K., Ledig, C., Newcombe, V.F., Simpson, J.P., Kane, A.D., Menon,
  D.K., Rueckert, D., Glocker, B.: Efficient multi-scale 3d cnn with fully
  connected crf for accurate brain lesion segmentation. Medical image analysis
  \textbf{36},  61--78 (2017)

\bibitem{kirillov2019panoptic}
Kirillov, A., Girshick, R., He, K., Doll{\'a}r, P.: Panoptic feature pyramid
  networks. In: Proceedings of the IEEE Conference on Computer Vision and
  Pattern Recognition. pp. 6399--6408 (2019)

\bibitem{lee2015deeply}
Lee, C.Y., Xie, S., Gallagher, P., Zhang, Z., Tu, Z.: Deeply-supervised nets.
  In: Artificial intelligence and statistics. pp. 562--570 (2015)

\bibitem{lin2017feature}
Lin, T.Y., Doll{\'a}r, P., Girshick, R., He, K., Hariharan, B., Belongie, S.:
  Feature pyramid networks for object detection. In: Proceedings of the IEEE
  conference on computer vision and pattern recognition. pp. 2117--2125 (2017)

\bibitem{lin2017focal}
Lin, T.Y., Goyal, P., Girshick, R., He, K., Doll{\'a}r, P.: Focal loss for
  dense object detection. In: Proceedings of the IEEE international conference
  on computer vision. pp. 2980--2988 (2017)

\bibitem{liu2015parsenet}
Liu, W., Rabinovich, A., Berg, A.C.: Parsenet: Looking wider to see better.
  arXiv preprint arXiv:1506.04579  (2015)

\bibitem{long2015fully}
Long, J., Shelhamer, E., Darrell, T.: Fully convolutional networks for semantic
  segmentation. In: Proceedings of the IEEE conference on computer vision and
  pattern recognition. pp. 3431--3440 (2015)

\bibitem{loshchilov2016sgdr}
Loshchilov, I., Hutter, F.: Sgdr: Stochastic gradient descent with warm
  restarts. arXiv preprint arXiv:1608.03983  (2016)

\bibitem{menze2014multimodal}
Menze, B.H., Jakab, A., Bauer, S., Kalpathy-Cramer, J., Farahani, K., Kirby,
  J., Burren, Y., Porz, N., Slotboom, J., Wiest, R., et~al.: The multimodal
  brain tumor image segmentation benchmark (brats). IEEE transactions on
  medical imaging  \textbf{34}(10),  1993--2024 (2014)

\bibitem{milletari2016v}
Milletari, F., Navab, N., Ahmadi, S.A.: V-net: Fully convolutional neural
  networks for volumetric medical image segmentation. In: 2016 fourth
  international conference on 3D vision (3DV). pp. 565--571. IEEE (2016)

\bibitem{myronenko20183d}
Myronenko, A.: 3d mri brain tumor segmentation using autoencoder
  regularization. In: International MICCAI Brainlesion Workshop. pp. 311--320.
  Springer (2018)

\bibitem{pereira2016brain}
Pereira, S., Pinto, A., Alves, V., Silva, C.A.: Brain tumor segmentation using
  convolutional neural networks in mri images. IEEE transactions on medical
  imaging  \textbf{35}(5),  1240--1251 (2016)

\bibitem{ronneberger2015unet}
Ronneberger, O., Fischer, P., Brox, T.: U-net: Convolutional networks for
  biomedical image segmentation. arXiv preprint arXiv:1505.04597  (2015)

\bibitem{shen2017boundary}
Shen, H., Wang, R., Zhang, J., McKenna, S.J.: Boundary-aware fully
  convolutional network for brain tumor segmentation. In: International
  Conference on Medical Image Computing and Computer-Assisted Intervention. pp.
  433--441. Springer (2017)

\bibitem{tan2019efficientdet}
Tan, M., Pang, R., Le, Q.V.: Efficientdet: Scalable and efficient object
  detection. arXiv preprint arXiv:1911.09070  (2019)

\bibitem{wu2018group}
Wu, Y., He, K.: Group normalization. In: Proceedings of the European conference
  on computer vision (ECCV). pp. 3--19 (2018)

\bibitem{zhao2018deep}
Zhao, X., Wu, Y., Song, G., Li, Z., Zhang, Y., Fan, Y.: A deep learning model
  integrating fcnns and crfs for brain tumor segmentation. Medical image
  analysis  \textbf{43},  98--111 (2018)

\bibitem{zhou2018unet++}
Zhou, Z., Siddiquee, M.M.R., Tajbakhsh, N., Liang, J.: Unet++: A nested u-net
  architecture for medical image segmentation. In: Deep Learning in Medical
  Image Analysis and Multimodal Learning for Clinical Decision Support, pp.
  3--11. Springer (2018)

\end{thebibliography}
% \bibitem{Vnet}
% F.Milletary, V-Net: Fully Convolutional Neural Networks for Volumetric Medical Image Segmentation
% \bibitem{focal}
% focal loss




% LNCS Homepage, \url{http://www.springer.com/lncs}. Last accessed 4
% \end{thebibliography}

\end{document}